\documentclass[10pt,a4paper,dvipdfmx]{article}
\usepackage{jcappub}
\usepackage{color}
\usepackage{amsmath,amssymb}

\newcommand{\YY}{ {\vec \Psi}}
\newcommand{\LL}{ {\cal L}}
\newcommand{\kk}{ {\vec k}}
\newcommand{\pp}{ {\vec p}}

\newcommand{\qq}{ {\vec q}}
\newcommand{\xx}{ {\vec x}}
\newcommand{\sr}{ {\vec s}}
\newcommand{\vv}{ {\vec v}}

\begin{document}

\title{Understanding redshift space distortions in density-weighted peculiar velocity}

\author[a,b,c]{Naonori S. Sugiyama,}
\author[c]{Teppei Okumura,}
\author[d]{and David N. Spergel}

\affiliation[a]{Department of Physics, School of Science, The University of Tokyo, Tokyo 113-0033, Japan}
\affiliation[b]{CREST, Japan Science and Technology Agency, Kawaguchi, Saitama, Japan}
\affiliation[c]{Kavli Institute for the Physics and Mathematics of the Universe (WPI),
		    Todai Institutes for Advanced Study, The University of Tokyo, Chiba 277-8582, Japan}
\affiliation[d]{Department of Astrophysical Sciences, Princeton University, Peyton Hall, Princeton NJ 08544-0010, USA}

\emailAdd{nao.s.sugiyama@gmail.com}

\abstract{

Observations of the kinetic Sunyaev-Zel'dovich (kSZ) effect measure the density-weighted
velocity field, a potentially powerful cosmological probe.
This paper presents an analytical method to predict 
the power spectrum and two-point correlation function of the density-weighted velocity in redshift space, the direct observables in kSZ surveys.  We show a simple relation between the density power spectrum and the density-weighted velocity power spectrum that  holds for both dark matter and halos.  Using this relation,
we can then extend familiar perturbation expansion techniques to the kSZ power spectrum.
One of the most important features of density-weighted velocity statistics in redshift space
is the change in sign of the cross-correlation between the density and density-weighted velocity
at mildly small scales due to nonlinear redshift space distortions.
Our model can explain this characteristic feature without any free parameters.
As a result, our results can precisely predict the non-linear behavior of the density-weighted velocity field  in redshift space
up to $\sim30\ h^{-1} {\rm Mpc}$ for dark matter particles at the redshifts of $z=0.0$, $0.5$, and $1.0$.
}


\arxivnumber{}

\notoc
\maketitle
\flushbottom

\section{Introduction}

Recent detections ~\cite{Hand:2012ui,Ade:2015lza}  of the  large-scale kinetic Sunyaev-Zel'dovich (kSZ) effect~\cite{Sunyaev:1980nv}, the Doppler effect due to the peculiar velocity of clusters relative to the CMB rest frame, open a new window into cosmology.  
These analyses measure the CMB signal at the position of galaxies in a spectroscopic survey, thus,
measure the relationship between the momentum field with the density field.  If the baryon physics is understood, these observations can directly test 
modified gravity theories and the effects of dark energy~\cite{Bhattacharya:2008,Ma:2013taq,Mueller:2014nsa} through measuring the relationship between density and velocity fields.
In the coming decade, we anticipate that new ground-based surveys such as Advanced ACT and space-based surveys such as Euclid and WFIRST will
make even more accurate measurements of the kSZ effect.
Therefore, predicting the precise non-linear behavior of the power spectrum and the two-point correlation function of the kSZ effect
in analytical approaches is an essential step in the interpretation of this data and in elucidating the nature of dark energy
and modified gravity theories.

This paper makes theoretical predictions for the
galaxy-weighted velocity field in redshift space, our observable when we  measure the CMB signal
at the position of galaxies identified in a redshift survey. 
The kSZ signal depends on the galaxy velocity field in two distinct ways:
 the kSZ signal measures the projected  electron momentum and
 the signal is measured at galaxy positions in redshift space.
Because we observe the signal in redshift space, we need to include
both 
 redshift space distortion (RSD)~\cite{Kaiser:1987qv} and finger-of-God effects in modeling
 the correlation function of the kSZ effect\cite{Okumura:2013zva}.

In section 2, we derive  a simple relation between the density power spectrum and the density-weighted velocity power spectrum, equation (\ref{pk_red}).
The relation holds for both of dark matter and halos and is one of the key results of the paper and enable us to use familiar perturbation theory results to link the calculations of the matter spectrum to measurements
of the kSZ power spectrum.  
Section 3 presents the analytical method to compute the density-weighted dark matter power spectrum
including the full non-linear effect of the redshift space distortion.
We apply the Lagrangian description and
expand the displacement vector from initial positions of particles in a perturbation expansion, 
because the Lagrangian description is derived from
the continuous limit of estimators of the power spectrum and the two-point correlation function.
Then, it is important to numerically compute
the power spectrum and two-point correlation function
with keeping the non-linear relation between them and the displacement vector in the Lagrangian description
\cite{Carlson:2012bu,Sugiyama:2013mpa,Vlah:2014nta},
enabling us to fully calculate the non-linear redshift space distortion effect 
through the coordinate transformation from real space to redshift space.
In Section 4,
we explain how to measure the density-weighted velocity power spectrum and two-point correlation function 
from $N$-body simulations.
Section 5 compares the analytical predictions and the results of the $N$-body simulations.
Section 6 concludes. 

Appendix~\ref{BAO} discuss the feature of the baryon acoustic oscillation in the two-point correlation function of the density-weighted velocity.
Appendix~\ref{HigherPole} discuss higher pole terms of the density-weighted velocity in the Legendre polynomial expansion, respectively.
In Appendix~\ref{ZA}, we compute the density-weighted velocity power spectrum and two-point correlation function 
in the Zel'dovich approximation from theory and measurements from particle distributions.
Appendix~\ref{SPT} compares difference between the Lagrangian and standard perturbation theories.
In Appendix~\ref{Volume}, 
in computing the correlation function
we investigate the impact of changing the minimum wavenumber in inverse Fourier transform of the power spectrum.
Appendix~\ref{Halos} shows the statistics for halos.

\section{Density-weighted velocity}
\label{2}

In real space, the radial component of the moments of the density-weighted velocity field, $p_{\parallel}^{(n)}$, are the
product of the density field, $\rho(\xx)$ and the comoving velocity, $\vv(\xx)$:
\begin{eqnarray}
	  p_{\parallel}^{(n)}(\xx) \equiv \left[ \hat{n}\cdot \vv(\xx) \right]^n \rho(\xx),
\end{eqnarray}
where $\hat{n}$ is the unit vector along the line of sight, and $\xx$ denotes the position in real space.

Our observations are in redshift space, $\sr$:
\begin{equation}
	\sr = \xx + \frac{\hat{n}\cdot\vv(\xx)}{aH}\hat{n},
	\label{Coord}
\end{equation}
 where $H$ is the Hubble parameter.
We can transform the density field through the continuity equation
$\rho(\sr)d^3s = \rho(\xx) d^3x$ and compute 
the density-weighted velocity  moments in redshift space:
\begin{eqnarray}
		p_{\parallel}^{(n)}(\sr) &=&  \int d^3x \left[ \hat{n}\cdot\vv(\xx) \right]^{n} \rho(\xx)
		\delta^{\rm (D)}\left(  \sr - \xx - \frac{\hat{n}\cdot\vv(\xx)}{aH}\hat{n} \right) \nonumber \\
		&=& \bar{\rho}\int d^3q \left[ a \hat{n}\cdot\dot{\YY}(\qq) \right]^n
		\delta^{\rm (D)}\left( \sr - \qq - \Psi_{\rm s}(\qq,\hat{n})\right) \nonumber \\
		&\to& \left( \frac{mN_{\rm p}}{a^3V} \right) \frac{N_{\rm mesh}^3}{N_{\rm p}} 
		\sum_{i=0}^{N_{\rm p}-1} \left[ \hat{n}\cdot\vv_i \right]^n \delta^{\rm (K)}\left( \sr - \sr_i \right) .
		\label{NV}
\end{eqnarray}
where the mean of mass density is given by the space-average with a survey volume $V$
\begin{eqnarray}
		\bar{\rho} \equiv \frac{1}{V} \int d^3x \rho(\xx) \to \left( \frac{mN_{\rm p}}{a^3V} \right).
\end{eqnarray}
The first, second, and third lines in Eq.~(\ref{NV}) 
are the Euler, Lagrangian, and particle descriptions, respectively.
The continuity equation $\rho(\xx)d^3x = \bar{\rho}d^3q$ relates the Eulerian and
Lagrangian descriptions as we transform to 
Lagrangian coordinates, $\qq$.
The right arrow denotes the discretization:
we use  $\int d^3q \to \left( V/N_{\rm p} \right)\sum_i$ and $\delta^{\rm (D)} \to \delta^{(K)}/V_{\rm mesh}$,
where $\delta^{\rm (D)}$ and $\delta^{\rm (K)}$ are the delta function and Kronecker delta,
and $V_{\rm mesh}$ is the volume in a shell we take.
Each physical quantity $a$, $\vv=a\dot{\YY}$, $m$, and $N_{\rm p}$
denotes the scale factor, the comoving velocity of particles,
mass of particles, and the number of particles.
The number of mesh $N_{\rm mesh}$ is defined as $N_{\rm mesh}^3 \equiv V/V_{\rm mesh}$.
The lowest moment, $n=0$, corresponds to the mass density $p_{\parallel}^{(n=0)} = \rho$.

In the Lagrangian description, the final positions of particles in real and redshift space 
are represented 
as 
\begin{eqnarray}
\xx &=& \qq + \YY(\qq) \nonumber \\ 
\sr &=& \qq + \YY_{\rm s}(\qq,\hat{n})\equiv \qq + \YY(\qq) + \frac{\hat{n}\cdot\dot{\YY}(\qq)}{H}\hat{n}
\end{eqnarray}
where  $\YY$  describes the displacement field and $\qq$ are the initial particle positions.

We define the perturbation of the density-weighted velocity with a general form, as
\begin{eqnarray}
		\delta p_{\parallel}^{(n)}(\sr) \equiv \frac{p_{\parallel}^{(n)}(\sr) - \bar{p}_{\parallel}^{(n)}}{\bar{\rho}} 
		&=&  \int d^3x \left[ \hat{n}\cdot\vv(\xx) \right]^{n} \frac{\rho(\xx)}{\bar{\rho}}
		\left[ \delta^{\rm (D)}\left(  \sr - \xx - \frac{\hat{n}\cdot\vv(\xx)}{aH}\hat{n} \right) - \frac{1}{V} \right] \nonumber \\
		&=& \int d^3q \left[a \hat{n}\cdot\dot{\YY}(\qq) \right]^n
		\left[ \delta^{\rm (D)}\left( \sr - \qq - \Psi_{\rm s}(\qq,\hat{n}) \right) -\frac{1}{V}  \right]\nonumber \\
		&\to& \frac{N_{\rm mesh}^3}{N_{\rm p}} 
		\sum_{i=0}^{N_{\rm p}-1} \left[ \hat{n}\cdot\vv_i \right]^n
		\left[ \delta^{\rm (K)}\left( \sr - \sr_i \right) - \frac{1}{N_{\rm mesh}^3} \right],
\end{eqnarray}
where the mean of the density-weighted velocity is
\begin{eqnarray}
		\bar{p}_{\parallel}^{(n)} &\equiv& \frac{1}{V} \int d^3s p_{\parallel}^{(n)}(\sr)
		= \frac{1}{V}\int d^3x \left[ \hat{n}\cdot\vv(\xx) \right]^n \rho(\xx) 
		= \frac{\bar{\rho}}{V} \int d^3q \left[a \hat{n}\cdot\dot{\YY}(\qq) \right]^n \nonumber \\
		&\to& \left( \frac{m N_{\rm p}}{a^3V} \right) \frac{1}{N_{\rm p}} \sum_{i=0}^{N_{\rm p}-1} \left[ \hat{n}\cdot\vv_i \right]^n.
\end{eqnarray}

The estimator of the two-point correlation function for the perturbation of the density-weighted velocity is given by
\begin{eqnarray}
	  \hat{\xi}_{\rm p}^{(n)(m)}(\sr) 
		&=& \frac{1}{V}\int d^3s_1 \delta p^{(n)}_{\parallel}(\sr+\sr_1) \delta p^{(m)}_{\parallel}(\sr_1), 
		\nonumber \\
		&=& 	\frac{a^{n+m}}{V} \int d^3q_1\int d^3q_2
		\left[ \hat{n}\cdot\dot{\YY}(\qq_1)  \right]^n \left[  \hat{n}\cdot\dot{\YY}(\qq_2) \right]^m \nonumber \\
		&\times&
		\left[ \delta^{\rm (D)}\left(\sr - (\qq_1-\qq_2) - \left( \YY_{\rm s}(\qq_1,\hat{n}) - \YY_{\rm s}(\qq_2,\hat{n}) \right) \right)
		- \frac{1}{V} \right]
		\nonumber \\
		&\to& \frac{N_{\rm mesh}^3}{N_{\rm p}^2} \sum_{i\neq j}
		\left[ \hat{n}\cdot\vv_i \right]^n\left[ \hat{n}\cdot\vv_j \right]^m 
		\delta^{\rm (K)}\left( \sr - \sr_{ij} \right) 
		+ \frac{N_{\rm mesh}^3}{N_{\rm p}^2} \sum_{i} \left[ \hat{n}\cdot\vv_i \right]^{n+m} \delta^{\rm (K)}(\sr) \nonumber \\
		&&
		 - \frac{1}{N_{\rm p}^2} \sum_{i, j}
		\left[ \hat{n}\cdot\vv_i \right]^n\left[ \hat{n}\cdot\vv_j \right]^m, 
		\label{xi1}
\end{eqnarray}
where $\sr_{ij} = \sr_i - \sr_j$ is the difference between final positions of particles.
Then, the two-point correlation function is given by
$\xi_{\rm p}^{(n)(m)} = \langle \hat{\xi}_{\rm p}^{(n)(m)}\rangle$, where $\langle \cdots \rangle$ means ensemble average.
This two-point correlation function by definition satisfies $\int d^3s \xi_{\rm p}^{(n)(m)}(\sr) = 0$. 
The second term in the final line behaves as the shot noise term in Fourier space.
In the case of $n = m = 0$,  
$\xi^{(0)(0)}_{\rm p} = \xi_{\rm m}$ is  the familiar density-density correlation function and the third term in the final line reduces to $-1$.
Furthermore, one can derive from Eq.~(\ref{xi1}) a theoretically convenient expression of 
the estimator of the two-point correlation function of the velocity field
\begin{eqnarray}
	  \hat{\xi}_{\rm p}^{(n)}(\sr)
	  &\equiv&  \sum_{m=0}^n \frac{(-1)^mn!}{m!\left( n-m \right)!}\hat{\xi}_{\rm p}^{(n-m)(m)}(\sr) \nonumber \\
		&=& \frac{N_{\rm mesh}^3}{N_{\rm p}^2} \sum_{i, j}
		\left[ \hat{n}\cdot\vv_i - \hat{n}\cdot\vv_j \right]^n 
		\left[  \delta^{\rm (K)}\left( \sr - \sr_{ij} \right)  - \frac{1}{N_{\rm mesh}^3} \right],
		\label{xi2}
\end{eqnarray}
where $\xi_{\rm p}^{(n)} = \langle\hat{\xi}_{\rm p}^{(n)}\rangle$. 
Note that the quantity measured in \cite{Hand:2012ui} is $\hat{\xi}_{\rm p}^{(1)}(\sr)$.
In this expression, the self-counting of particles  vanishes
due to the weight function $\left[ \hat{n}\cdot\vv_i - \hat{n}\cdot\vv_j \right]^n$.
The goal of this paper is to analytically predict this expression 
and compare with the results from $N$-body simulations.

The power spectrum of the density-weighted velocity is defined as the Fourier transformation of
the two-point correlation functions $P_{\rm p}^{(n)(m)} \equiv \int d^3s e^{-i\kk\cdot\sr} \xi^{(n)(m)}(\sr)$
and $P_{\rm p}^{(n)} \equiv \int d^3s e^{-i\kk\cdot\sr} \xi^{(n)}(\sr)$.  Thus,
the power spectrum estimators,
\begin{eqnarray}
	  \hat{P}^{(n)(m)}_{\rm p}(\kk) 
		&=&  \frac{V}{N_{\rm p}^2} 
		\sum_{i, j} \left[ \hat{n}\cdot\vv_i \right]^n \left[ \hat{n}\cdot\vv_j \right]^m
		\left[ e^{-i\kk\cdot\sr_{ij}} - \delta^{\rm (K)}(\kk) \right] \nonumber \\
	  \hat{P}^{(n)}_{\rm p}(\kk) 
		&=&  \frac{V}{N_{\rm p}^2} \sum_{i, j} \left[ \hat{n}\cdot\vv_i - \hat{n}\cdot\vv_j\right]^n 	
		\left[ e^{-i\kk\cdot\sr_{ij}} - \delta^{\rm (K)}(\kk) \right],
		\label{pk}
\end{eqnarray}
are directly related to the underlying power spectra, $P^{(n)(m)}_{\rm p} = \langle  \hat{P}^{(n)(m)}_{\rm p} \rangle$
and $P^{(n)}_{\rm p} = \langle\hat{P}^{(n)}_{\rm p}\rangle$.
These power spectrum can also be measured from data by correlating a CMB map with the $p$-th power of the reconstructed velocity field.
In particular, the estimators of the power spectra $\hat{P}_{\rm p}^{(n\geq 1)}$ defined here have no shot-noise term,
because the self-counting of particles are removed due to the weight function $\left( \hat{n}\cdot\vv_i - \hat{n}\cdot\vv_j \right)^n$.
We can ignore the second term proportional to $\delta^{\rm (K)}(\kk)$ in measuring the power spectrum,
because the term only contributes to a bin including $\kk=0$ and guarantees $P^{(n)(m)}(\kk=0) = P^{(n)}(\kk=0)=0$
which corresponds to $\int d^3s \xi^{(n)(m)}(\sr) = \int d^3s \xi^{(n)}(\sr) = 0$.
Therefore, we only have to set $P^{(n)(m)}(\kk=0) = P^{(n)}(\kk=0)=0$ at the end of calculation by hand.

Since the power spectra are not isotropic, we expand the density-weighted power spectrum and two-point correlation function in the Legendre polynomials.
\begin{eqnarray}
	  P_{\rm p}^{(n)}(\kk,\hat{n}) = 
	  \sum_{\ell=0}^{\infty}P_{\rm p,\ell}^{(n)}(k) {\cal L}_{\ell}(\hat{k}\cdot\hat{n}), \quad
	  {\rm and} \quad
	  \xi_{\rm p}^{(n)}(\sr,\hat{n}) = 
	  \sum_{\ell=0}^{\infty}\xi_{\rm p,\ell}^{(n)}(s) {\cal L}_{\ell}(\hat{s}\cdot\hat{n}),
 \end{eqnarray}
where the two-point correlation function is given by
\begin{eqnarray}
	  \xi_{\rm p,\ell}^{(n)}(s) = i^{\ell}\int\frac{d^3k}{(2\pi)^3} j_{\ell}(sk) P^{(n)}_{\rm p,\ell}(k).
	  \label{pk_to_xi}
\end{eqnarray}
Note that $\xi_{\rm p}^{(n = \rm odd)}$ and $\xi_{\rm p}^{(n=\rm even)}$
contain only odd- and even-pole terms in the Legendre expansion, respectively,
and the same is true for the power spectrum.

Eq.~(\ref{pk}) yields a simple relation between the estimators of the power spectra for density-weighted velocity and dark matter particles
 \begin{eqnarray}
	   \hat{P}_{\rm p}^{(n)(m)}(\kk) &=&  (-1)^m \left( i \frac{aH}{\kk\cdot\hat{n}} \right)^{n+m} 
	   \left[\frac{d^n}{d \gamma_1^n}\frac{d^m}{d \gamma_2^m} \hat{P}_{\rm p}^{(0)(0)}\left( \kk;\gamma_1,\gamma_2 \right)  \right]
		\bigg|_{\gamma_1 = \gamma_2 = 1}, \nonumber \\
		\hat{P}_{\rm p}^{(n)}(\kk) &=&  
		\left( i\frac{aH}{\kk\cdot\hat{n}} \right)^n\left[ \frac{d^n}{d \gamma^n} \hat{P}^{(0)}_{\rm p}(\kk;\gamma) \right]\bigg|_{\gamma = 1}.
		\label{pk2}
\end{eqnarray}
where the spectra of the lowest moment velocity are defined as
\begin{eqnarray}
	  \hat{P}_{\rm p}^{(0)(0)}\left( \kk;\gamma_1,\gamma_2 \right)
		&\equiv& 
		\frac{V}{N_{\rm p}^2}
		\sum_{i, j}	
		\left[ e^{-i\kk\cdot\xx_{ij} - i\frac{\kk\cdot\hat{n}}{aH}\left( \gamma_1 \hat{n}\cdot\vv_i - \gamma_2 \hat{n}\cdot\vv_j \right)} \right]
		,\nonumber \\
	\hat{P}_{\rm p}^{(0)}\left( \kk;\gamma\right) 
		&\equiv& 
		\frac{V}{N_{\rm p}^2}
		\sum_{i, j}
		\left[ e^{-i\kk\cdot\xx_{ij} - i\gamma \frac{\kk\cdot\hat{n}}{aH}\left( \hat{n}\cdot\vv_i - \hat{n}\cdot\vv_j \right)} \right].
\end{eqnarray}
Note that $\hat{P}^{(n={\rm odd})}_{\rm p}$ are by definition imaginary.
In the case that $\gamma_1$, $\gamma_2$, and $\gamma$ are unity, $\hat{P}_{\rm p}^{(0)(0)}(\kk,\gamma_1=1,\gamma_2=1)$ and $\hat{P}_{\rm p}^{(0)(0)}(\kk,\gamma=1)$
reduce to the matter power spectrum $\hat{P}_{\rm m}(\kk) = \hat{P}_{\rm p}^{(0)(0)}(\kk) = \hat{P}_{\rm p}^{(0)}(\kk)$ in redshift space.
Thus, the matter power spectrum behaves as the generating function of the density-weighted velocity power spectrum 
~\cite{Scoccimarro:2004}.
The expressions of $\hat{P}_{\rm p}^{(n)(m)}$
are equivalent to those used in the distribution function approach method
~\cite{Seljak:2011tx, Okumura:2011pb, Okumura:2012xh, Vlah:2012ni, Vlah:2013lia, Okumura:2013zva}.
In particular, provided that the velocity field $\vv$ is proportional to $f= d \ln D/ d\ln a$
with $D$ being the linear growth factor,
we finally derive from Eq.~(\ref{pk2})
\footnote{
In real space, this expression reduces to
\begin{eqnarray}
	  \hat{P}_{\rm p}^{(n)}(\kk)
	  =  \left( i\frac{aHf}{\kk\cdot\hat{n}} \right)^n \left[ \frac{\partial^n}{\partial f^n} \hat{P}_{\rm m}(D,f,\kk,\hat{n})  \right] \Bigg|_{f=0}.
		\label{pk_real}
\end{eqnarray}
}
\begin{eqnarray}
	  \hat{P}_{\rm p}^{(n)}(\kk,\hat{n})
		=  \left( i\frac{aHf}{\kk\cdot\hat{n}} \right)^n\frac{\partial^n}{\partial f^n}
    	\hat{P}_{\rm m}(D,f,\kk,\hat{n}).
		\label{pk_red}
\end{eqnarray}
This is the main result in this paper.
 Eq.~(\ref{pk_red}) relates all of the power spectra of density-weighted velocity  moments in redshift space to the matter power spectrum including redshift space distortions. 
Note that this expression is derived from the particle description so can be used 
as the estimators for measuring the power spectra.
Therefore, Eq.~(\ref{pk_red})
is valid as long as  the electron velocity is proportional to the growth rate  of structure, $\vv \propto f$.
This expression holds even for halos as far as the velocities of halos are defined as 
a linear combination of velocities of dark matter particles such as velocities of halo centers, 
because then the velocities of halos are still proportional to the growth rate $f$,
and the power spectrum for halos is measured using the same estimator as that for dark matter particles.
In the following sections,
we compute the power spectrum and two-point correlation function 
of the density-weighted velocity for dark matter particles
using Eq.~(\ref{pk_red}) and Lagrangian perturbation techniques, and compare with the results measured from $N$-body simulations.
If we had used alternative approaches such as 
the renormalized perturbation theory~\cite{Crocce:2006},
the effective field theory approach~\cite{Baumann:2010tm,Carrasco:2012cv,Porto:2013qua},
the convolution Lagrangian perturbation theory~\cite{Carlson:2012bu},
the distribution function approach~\cite{Seljak:2011tx, Okumura:2011pb, Okumura:2012xh, Vlah:2012ni, Vlah:2013lia, Okumura:2013zva},
the integrated perturbation theory~\cite{Matsubara:2007wj, Matsubara:2008wx, Matsubara:2013ofa},
and the TNS model~\cite{Taruya:2010mx}, to compute the density power spectrum, Eq.~(\ref{pk_red}) could still be used
to compute the density-weight velocity power spectrum.

\section{Theoretical calculations}
\label{TheoreticalModel}

\subsection{Analytical expressions}
We have shown that the power spectrum of density-weighted velocity 
can be directly derived from the matter density power spectrum using Eq.~(\ref{pk_red}).
The analytical expression of the power spectrum is given by using the ensemble average $\langle \cdots \rangle$ for Eq.~(\ref{pk})
\begin{eqnarray}
	  \left\langle \hat{P}_{\rm m}(\kk) \right\rangle
		&=& \frac{V}{N_{\rm p}^2}
		\sum_{j=0}^{N_{\rm p}-1} \sum_{|i-j|=0}^{N_{\rm p}-1}
		e^{-i\kk\cdot\left( \qq_i-\qq_j \right)}
		\left\langle e^{-i\kk\cdot\left( \YY_{\rm s}(\qq_i-\qq_j,\hat{n}) - \YY_{\rm s}(0,\hat{n}) \right)} \right\rangle \nonumber \\
		&=&  \frac{V}{ N_{\rm p} } \sum_{\alpha=0}^{N_{\rm p}-1}
		e^{-i\kk\cdot\qq_{\alpha}}
		\left\langle e^{-i\kk\cdot\left( \YY_{\rm s}(\qq_{\alpha},\hat{n}) - \YY_{\rm s}(0,\hat{n}) \right)} \right\rangle \nonumber \\
		&=&  \frac{V}{ N_{\rm p} } \sum_{\alpha=0}^{N_{\rm p}-1}
		e^{-i\kk\cdot\qq_{\alpha}}e^{\Sigma(\kk,\qq_{\alpha},\hat{n})-\bar{\Sigma}(\kk,\hat{n})},
		\label{pk_an}
\end{eqnarray}
where we used $\sum_{i, j} = \sum_{j=0}^{N_{\rm p}-1} \sum_{|i-j|=0}^{N_{\rm p}-1}$ and $\alpha = |i-j|$,
and we expressed the final positions of particles using the displacement vector including the redshift space distortion:
$\sr_i = \qq_i + \YY(\qq_i) + \frac{\hat{n}\cdot\dot{\YY}(\qq_i)}{H}\hat{n} \equiv
\qq_i + \YY_{\rm s}(\qq_i,\hat{n})$.
Note that in a context of simulations the ensemble average means averaging measured power spectra using infinite realizations.
From the translation symmetry of the ensemble average in the first line of Eq.~(\ref{pk_an}), 
we can take $j=0$ in the summation $\sum_{|i-j|=0}^{N_{\rm p}-1}$ without loss of generality,
remaining the single summation $\sum_{\alpha=0}^{N_{\rm p}-1}$ in the second line.
Furthermore, 
we need to set $P^{(n)}(\kk=0)=0$ at the end of calculation, because we ignored the delta function $\delta^{\rm (D)}(\kk)$ in Eq.~(\ref{pk}).
The correlation functions of the displacement vector are defined 
using the cumulant $\langle \cdots \rangle_{\rm c}$ as~\cite{Matsubara:2007wj,Sugiyama:2013mpa}
\begin{eqnarray}
		\Sigma(\kk,\qq_{\alpha},\hat{n}) &=&  \sum_{n=2}^{\infty}\sum_{m=1}^{n-1} \frac{(-i)^n(-1)^m}{m!(n-m)!}
		\left\langle \left[ \kk\cdot\YY_{\rm s}(\qq_{\alpha},\hat{n}) \right]^{n-m}
		\left[ \kk\cdot\YY_{\rm s}(0,\hat{n}) \right]^m  \right\rangle_{\rm c},
		\nonumber \\
		\bar{\Sigma}(\kk,\hat{n}) &=& \Sigma(\kk,\qq_{\alpha}=0,\hat{n})
		= -2\sum_{n=1}^{\infty}\frac{(-1)^n}{(2n)!}	
		\left\langle \left[ \kk\cdot\YY_{\rm s}(0,\hat{n})  \right]^{2n}\right\rangle_{\rm c}.
		\label{sigma}
\end{eqnarray}
We can directly compute Eq.~(\ref{pk_an}) using the Discrete Fourier Transformation (DFT),
even though the computational cost is expensive:
for a wavenumber $\kk={\cal O}\left( N_{\rm mesh}^3 \right)$, we need the summation of the number of particles ${\cal O}\left( N_{\rm p} \right)$,
resulting in the computational cost ${\cal O}\left( N_{\rm mesh}^3 \times N_{\rm p} \right)$.

\subsection{Approximation method}
We present a technique to quickly compute the matter power spectrum in Eq.~(\ref{pk_an}) as studied in~\cite{Sugiyama:2013mpa}.
First, we express the summation in Eq.~(\ref{pk_an}) using an integral representation
$\frac{V}{N_{\rm p}}\sum_{\alpha=0}^{N_{\rm p}-1} = \int d^3q$.
Next, we expand the correlation function of the displacement vector $\Sigma$ in the associated Legendre polynomials
\begin{eqnarray}
		\Sigma(\kk,\qq, \hat{n})
		= \sum_{\ell=0}^{\infty} \sum_{m=0}^{\ell} i^{\ell}
		\Sigma_{\ell}^m(k,q,\mu_k) \LL_{\ell}^m(\mu) \cos\left( m\varphi \right),
		\label{Sigma}
\end{eqnarray}
where $\mu = \hat{k}\cdot\hat{q}$, $\mu_{k} = \hat{n}\cdot\hat{k}$, and $\hat{n}\cdot\hat{q} = \mu \mu_k + \sqrt{1-\mu^2}\sqrt{1-\mu_k^2}\cos(\varphi)$.
Finally, we use the following expansion for the power spectrum
\begin{eqnarray}
	  \left\langle \hat{P}_{\rm m}(\kk)\right \rangle 
		=  \sum_{n=0}^{\infty}\frac{1}{n!} \int dq q^2 d\mu d\varphi
		e^{-ikq \mu}
		e^{\Sigma_0^0(k,q,\mu_k)-\bar{\Sigma}(k,\mu_k)}
		\left( \sum_{\ell=1}^{\infty}\sum_{m=0}^{\ell} i^{\ell} \Sigma_{\ell}^{m}(k,q,\mu_k)
		\LL_{\ell}^{m}(\mu) \cos\left( m\varphi \right) \right)^n.
		\label{Pkan}
\end{eqnarray}
By analytical calculation of the angular integral in $\int d^3q = \int dq q^2 \int d\mu d\varphi$,
the 3-dimensional integral reduces to the single integral $\int dq q^2$.
In our previous work~\cite{Sugiyama:2013mpa}, we verified that this expression quickly converges to the true Lagrangian power spectrum. In this paper we compute this expansion up to $n=2$. 

It is worth noting that the expression in Eq.~(\ref{Pkan}) includes the volume integral of the Lagrangian coordinate $\int d^3q$,
which comes from the summation of pairs of particles in the estimator of the power spectrum as shown in Eq.~(\ref{pk_an})
\footnote{
As a simple relation to previous works, 
the $\Gamma$-expansion~\cite{Bernardeau:2008fa} is the general formula to classify the non-linear correction terms to the matter power spectrum 
with mode-coupling integrals in Fourier space.
The power spectrum in the $\Gamma$-expansion is given by
\begin{eqnarray}
		P_{\rm m}(k) = \sum_{n=1}^{\infty}
		\int \frac{d^3p_1}{(2\pi)^3} \dots \int \frac{d^3p_n}{(2\pi)^3} \left( 2\pi \right)^3 \delta^{\rm (D)}(\kk-\pp_1-\dots-\pp_n)
		\left[ \Gamma^{(n)}(\pp_1,\dots,\pp_n) \right]^2 P_0(p_1) \cdots P_0(p_n),\nonumber
\end{eqnarray}
where $\pp_1 \dots \pp_n$ are wavenumber and $P_0$ is the initial linear power spectrum.
Since infinite mode-coupling integrals are generated from the non-linear relation between
the power spectrum and the displacement vector in Eq.~(\ref{pk_an}),
through Eq.~(\ref{Pkan}) we can compute infinite orders of the $\Gamma$-expansion (see Sec.5.3. in \cite{Sugiyama:2013mpa}).
}.
Furthermore, we keep $\exp\left( \Sigma_0^0(k,q,\mu_k)  \right)$ in the integrand of Eq.~(\ref{Pkan}) un-expanded.
The expression in Eq.~(\ref{Pkan}) contains infinite mode-couping integrals in Fourier space.
Expanding $\exp\left( \Sigma \right)$ and truncating it at a finite order,
$\exp\left( \Sigma \right) = 1 + \Sigma + \frac{1}{2}(\Sigma)^2+\dots$, yields finite
mode-coupling integrals through the convolution theorem.
These facts do not depend on the functional form of the correlation function of the displacement vector in Eq.~(\ref{Sigma}),
and even the Zel'dovich approximation~\cite{Zeldovich:1969sb}, which is the linear approximation of the displacement vector,
contains infinite mode-couping integrals~\cite{Crocce:2006,Sugiyama:2013mpa}.

As pointed out in~\cite{Tassev:2013rta,White:2014gfa},
the Zel'dovich approximation can well explain the broadening and evolution of the baryon acoustic peak in the density correlation function
around the scales of $r\gtrsim 60h^{-1}{\rm Mpc}$.
Furthermore, many resummation techniques of bulk flow motions in the Eulerian description of perturbation theories
to accounts for the non-linear feature of the baryon acoustic peak
are closely related to the Zel'dovich approximation (e.g., see Introduction in the original renormalized perturbation theory paper~\cite{Crocce:2006}).
The positions and velocities of dark matter particles in the Zel'dovich approximation
evolve in the linear gravitational potential through the equation of motion in Eq.~(\ref{EOM}),
while the density field evolves non-linearly through the relation between the density field and the position of particles in Eq.~(\ref{NV}).
In this sense, the main contributions to
the broadening of the baryon acoustic peak comes from the change in particle positions in the linear gravitational potential,
and non-linear gravitational effects are negligible.
In Appendix~\ref{BAO},
we demonstrate that the Zel'dovich approximation explains the behavior of the peak broadening 
even in the density-weighted velocity correlation function.

The full non-linear continuity equation plays a key role to explain the effect of the redshift space distortion, 
because the effect comes only from the coordinate transformation via the continuity equation $\bar{\rho}d^3q = \rho(\xx) d^3x = \rho(\sr) d^3s$. 
For analytic prediction of the redshift space distortion effect,
we need to compute infinite mode-coupling integrals from the velocity field through the continuity equation,
which is achieved by the volume integral in Eq.~(\ref{Pkan}).

\subsection{Gravitational effects}

The equation-of-motion describes  the non-linear gravitational effects on the evolution of the 
displacement vector:
\begin{eqnarray}
		\ddot{\YY}(\qq_i) + 2H \dot{\YY}(\qq_i) = - \frac{1}{a^2} \frac{\partial}{\partial \xx} \delta \phi(\xx)
		\label{EOM}
\end{eqnarray}
with the irrotational condition $\nabla \times \vv = 0$
and the Poisson equation $\frac{\nabla^2}{a^2} \delta \phi = \frac{3}{2}H^2\Omega_{\rm m}\delta_{\rm m}$, 
where $\delta \phi$ is the gravitational potential, 
and $\Omega_{\rm m}$ is the cosmological parameter of total matter.
We expand this equation in a perturbation expansion.
The displacement vector is described in the perturbation expansion using $D$ and $f$,
\begin{eqnarray}
		\YY(z,\qq) = \sum_{\alpha = 1}^{\infty} D^{\alpha}(z) \YY^{(\alpha)}(z=0,\qq),
\end{eqnarray}
where $\YY^{(\alpha)} = {\cal O}\left( (\delta_{\rm lin})^{\alpha} \right)$
with $\delta_{\rm lin}$ being the linear matter density perturbation.
This separation of variables holds in the linearized theory or in the case of $f^2 = \Omega_{\rm m}$ in the non-linear theory.
Since it is well known that $f \approx \Omega_{\rm m}^{0.55}$ is a good approximation in General relativity (GR)~\cite{Linder:2005in},
we use this expansion for any cosmology in GR.
Then, the velocity field $\vv = \dot{\YY}$ is proportional to $f$, 
and we can use the simple relation between the density-weighted power spectrum
and the matter power spectrum in Eq.~(\ref{pk_red}).

In the linearized theory, we derive from Eq.~(\ref{pk_red})
\begin{eqnarray}
	  \left\langle \hat{P}_{\rm p}^{(1)}(\kk) \right\rangle
		&=&   \left( i\frac{aHf}{k \mu_k} \right)\frac{\partial}{\partial f} \left( b + f\mu_k^2 \right)^2 D^2 P_{0}(k) 
		=  2i aH f  \mu_k \left( b + f\mu_k^2 \right) D^2 \frac{P_{\rm 0}(k)}{k}, \nonumber\\
		\left \langle \hat{P}_{\rm p}^{(2)}(\kk) \right\rangle
		&=&   \left( i\frac{aHf}{k \mu_k} \right)^2\frac{\partial^2}{\partial f^2}
		\left( b + f\mu_k^2 \right)^2 D^2 P_{0}(k) 
		=  -2 \left( aHf \right)^2  \mu_k^2 D^2 \frac{P_{\rm 0}(k)}{k^2}, 
\end{eqnarray}
where $P_{\rm 0}$ denotes the linear power spectrum at the present time,
and $\left( b + f\mu_k^2 \right)^2$ is the Kaiser factor~\cite{Kaiser:1987qv} with the linear halo bias $b$.
It should be noted that the difference between $P^{(1)(1)}$ and $P^{(2)}$ in Eq.~(\ref{pk})
comes from non-linear corrections, and $P^{(1)(1)}$ is simply related to $P^{(2)}$ in linear theory: $P^{(1)(1)} = - P^{(2)}/2$.
These expressions are consistent with the result of the previous work (Eq.~(2.22) in~\cite{Okumura:2013zva}).

We consider up to the third order displacement vector in the perturbation theory: $\YY = \YY^{(1)} + \YY^{(2)} + \YY^{(3)}$.
Then, the correlation functions of the displacement vectors $\Sigma$ are truncated up to the 1-loop order
$\Sigma = \Sigma_{\rm lin} +\Sigma_{\rm 1\mathchar`-loop}  + {\cal O}\left( (P_0)^3 \right)$,
where ``$n$-loop'' means ${\cal O}\left( P_0^{n+1} \right)$.
Specific expressions of $\Sigma$ at the 1-loop order are summarized in~\cite{Sugiyama:2013mpa}.

\subsection{Finite volume and sampling effects in simulations}

To account for the finite volume and finite particle effects in simulations,
we restrict the range of wavenumber in the linearized power spectrum we use for calculation:
$2\pi/{\rm L} < k < 2\pi N_{\rm p}^{1/3}/{\rm L}$, where $L$ denotes box size in simulations.
Furthermore, we choose the range of the volume integration as $0 \leq q \leq {\rm L}$ in Eq.~(\ref{pk_an}).

\section{Measurement of the power spectrum and two-point correlation function from $N$-body simulations}

Unfortunately, evaluating Equation (\ref{pk}) is computationally expensive:  computing
 a vector of wavenumber $\kk = {\cal O}\left( N_{\rm mesh}^3 \right)$
requires  ${\cal O}\left(N_{\rm mesh}^3 \times N_{\rm p} \right)$ operations.
To speed the calculation, 
we define an alternative estimator of the power spectrum:  the square of the Fourier-transformed density-weighted velocity field given by Eq.~(\ref{NV})
using the cloud in cell (CIC) particle assignment method with the finite number of grid points $N_{\rm mesh}$,
\begin{eqnarray}
	  \hat{P}^{(n)(m)}_{\rm p}(\kk) &=&
		\left[ \delta p_{\parallel}^{(n)}(\kk)  \right]\left[ \delta p_{\parallel}^{(m)}(\kk) \right]^*
		\nonumber \\
		\hat{P}^{(n)}_{\rm p}(\kk) &=&
		\sum_{m=0}^{n} \frac{(-1)^m n!}{m!\left( n-m \right)!}
		\left[ \delta p_{\parallel}^{(n-m)}(\kk)  \right]\left[ \delta p_{\parallel}^{(m)}(\kk) \right]^* 
		\label{pk_measure}
\end{eqnarray}
where $\delta p_{\parallel}^{(n)}(\kk) = \int d^3s e^{-i\kk\cdot\sr} \delta p_{\parallel}^{(n)}(\sr)$.
We adopt $N_{\rm mesh}=512$, and in the limit of large $N_{\rm mesh}$, Eq.~(\ref{pk_measure}) converges to Eq.~(\ref{pk}).
Here, the shot noise term in $\hat{P}_{\rm p}^{(n)(m)}$ is given by
\begin{eqnarray}
	\frac{C_{\rm CIC}(\kk)}{\bar{n}} 
	\left( \frac{1}{N_{\rm p}} \sum_{i=0}^{N_{\rm p}-1} \left[ \hat{n}\cdot\vv_i \right]^{n+m}  \right),
\end{eqnarray}
where $\bar{n} \equiv N_{\rm p}/V$, and the proportional factor in the shot noise term is derived from
the second term in the final line of Eq.~(\ref{xi1}),
while $\hat{P}_{\rm p}^{(n\geq 1)}$ has no shot-noise term 
due to the cancellation of the self-counting of particle pairs
in the weight function $\left( \hat{n}\cdot\vv_i - \hat{n}\cdot\vv_j \right)^n$ as mentioned in Sec.~\ref{2}.
The scale-dependent correction function to the shot-noise term for the CIC particle alignment $C_{\rm CIC}$ is given by~\cite{Jing:2004fq}
\begin{eqnarray}
		C_{\rm CIC}(\kk) = \prod_{i=x,y,z}\left[ 1 -\frac{2}{3}\sin^2\left(\frac{\pi k_i}{2 k_{\rm N}}\right) \right],
\end{eqnarray}
where $k_{\rm N} = \pi N_{\rm mesh}/L$ with $L$ being box size of simulations.
These estimators satisfy $\hat{P}^{(n)(m)}(\kk=0) = \hat{P}^{(n)}(\kk=0) = 0$.

We compute the two-point correlation function as the inverse Fourier transformation 
$\xi^{(n)(m)}(\sr) = \int \frac{d^3k}{(2\pi)^3}e^{i\kk\cdot\sr} P^{(n)(m)}(\kk)$
and $\xi^{(n)}(\sr) = \int \frac{d^3k}{(2\pi)^3}e^{i\kk\cdot\sr} P^{(n)}(\kk)$,
because the analytical approach also use the inverse Fourier transformation 
for computing the two-point correlation function in Eq.~(\ref{pk_to_xi}).
The computed two-point correlation function by definition satisfies
$\int d^3s \xi^{(n)(m)}(\sr) = \int d^3s \xi^{(n)}(\sr) = 0$ due to the condition of $P^{(n)(m)}(\kk=0) = P^{(n)}(\kk=0) = 0$.
In inverse Fourier transform to obtain the power spectrum,
we use exactly the same integral range of wavenumber
for the analytical prediction and the measurement from simulation data for fair comparison.
We have checked that our simulation results for the correlation functions agree with the results using one double the Nyquist frequency 
within a few percent level at scales larger than $\sim 20$ $h^{-1}{\rm Mpc}$.
This is sufficient for our purpose, which is to compare perturbation theories with $N$-body simulations for the density-weighted velocity statistics
at mildly non-linear scales.

\section{Results}

\begin{figure}[t]
	  \begin{center}
	  \psfig{figure=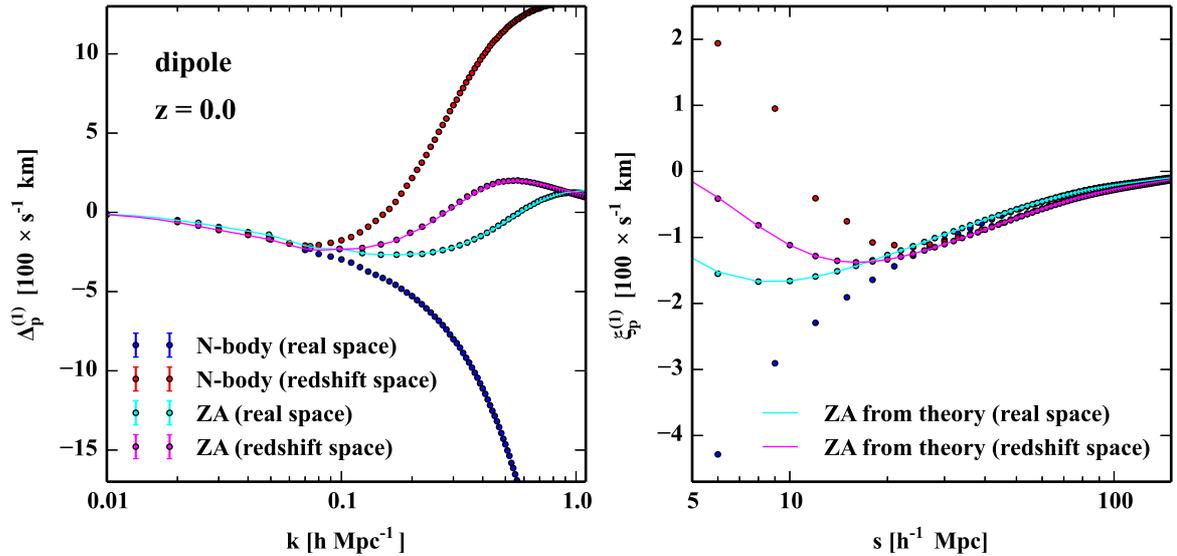}
		\end{center}
		\vspace{-0.5cm}
		\caption{
		Dipole terms of the density-weighted velocity power spectrum $\Delta^{(1)}_{\rm p}$
		and two-point correlation function $\xi^{(1)}_{\rm p}$
		are plotted in both real and redshift space at a redshift of $z=0$. 	
		Blue and red symbols denote the results of the $N$-body simulations in real and redshift space,
		and cyan and magenta symbols are measured from the particle distribution in the Zel'dovich approximation.
		Cyan and magenta solid lines represent the numerically computed analitical predictions in Eqs.~(\ref{pk_red}) and (\ref{Pkan}).
		The quantity that was measured in~\cite{Hand:2012ui}
		is the two-point correlation function of the galaxy halos which is closely
		related to the matter two-point correlation function $\xi^{(1)}_{\rm p}$ in redshift space.
		If two particles are moving toward each other,
		their contribution to $\Delta_{\ell=1}^{(1)}$ and $\xi_{\ell=1}^{(1)}$ will be negative, 
		$\Delta_{\ell=1}^{(1)}<0$ and $\xi_{\ell=1}^{(1)}<0$,
		and if moving apart, positive $\Delta_{\ell=1}^{(1)}>0$ and $\xi_{\ell=1}^{(1)}>0$.
		Gravitational attraction predicts a slight tendency of any pair of objects 
		to be moving toward rather than away from each other at large scales, 
		resulting in the negative values of $\Delta_{\ell=1}^{(1)}$ and $\xi_{\ell=1}^{(1)}$.
		The signs of $\Delta_{\ell=1}^{(1)}$ and $\xi_{\ell=1}^{(1)}$ change from negarive to positive at mildly non-linear scales
		in both the $N$-body simulation and Zel'dovich approximation,
		because the positions of particles move in the direction of their line-of-sight velocity through the coordinate transformation
		from real to redshift space, and
		different two particles apparently path through and move away from each other in redshift space.
    		}
		\label{fig1}
\end{figure}

For the $N$-body simulations,
we use $Gadget2$~\cite{Springel:2000yr,Springel:2005mi}
with the initial condition in the Zel'dovich approximation
generated by $2LPT$~\cite{Crocce:2006ve} at a redshift of $z=99.0$.
We use the best fitting cosmological parameters from $Planck2015$~\cite{Planck:2015xua}:
$\Omega_{\rm m} = 0.308$, $\Omega_{\rm \Lambda} = 0.692$,
$\Omega_{\rm b} = 0.048$, $h = 0.678$, $\sigma_8 = 0.815$, and $n_s = 0.968$.
We calculate the linearized power spectrum using $CLASS$~\cite{Lesgourgues:2011re}
The box size $L$ and the number of particles $N_{\rm p}$ are
$L = 1024\ [h^{-1}\ {\rm Mpc}]$ and $N_{\rm p} = 512^3$, respectively.
The number of realizations is 30.
The errors in simulation are estimated by $\sqrt{Cov}\left( P_{\rm p,\ell}^{(n)}(k),P_{\rm p,\ell}^{(n)}(k) \right)/\sqrt{30}$
and
$\sqrt{Cov}\left( \xi_{\rm p,\ell}^{(n)}(s),\xi_{\rm p,\ell}^{(n)}(s) \right)/\sqrt{30}$,
which are typically smaller than the size of circles in figures, and therefore, are hardly visible.
We set bin widths in the power spectrum and correlation function 
as $dk = 0.01 h{\rm Mpc}^{-1}$ and $ds = 3h^{-1}{\rm Mpc}$, respectively.
Note that since the purpose of this paper is to test the accuracy of our new formula (Eq.\ref{pk_red}), we do not consider any observational systematics such as astrophysical effects on kSZ surveys. More realistic forecasts will be provided in our future paper (Sugiyama, Okumura, Spegel in prep.).

We compare our analytical predictions from Eqs.~(\ref{Pkan}) and (\ref{pk2}) with the $N$-body simulation results.
In doing so, we define
\begin{eqnarray}
	  \Delta_{\rm p, \ell}^{(n)}(\kk) \equiv i^{\ell}\frac{k^3}{2\pi^2} P_{\rm p, \ell}^{(n)}(\kk),
	  \label{Delta}
\end{eqnarray}
where $\Delta_{\rm p, \ell}^{(0)}$ is the dimensionless power spectrum of density fluctuations, and 
$\Delta_{\rm p, \ell}^{(n\geq 1)}$ have the dimension of the $n$th power of velocity $[{\rm km/s}]^n$.
It should be noted that the quantity that was measured in~\cite{Hand:2012ui} is closely 
related to the dipole of the matter two-point correlation function $\xi^{(1)}_{\rm p, 1}$ in redshift space.
Therefore, we mainly focus on $\Delta_{\ell=1}^{(1)}$ and $\xi_{\ell=1}^{(1)}$ in this paper.
Furthermore, as one of applications of our formalism in Eq.~(\ref{pk_red})
we investigate higher order density-weighted velocity statistics $P_{\rm p}^{(n=2)}$ and $\xi_{\rm p}^{(n=2)}$.
The properties of higher multi-pole moments of $P^{(1)}_{\rm p}$, $\xi^{(1)}_{\rm p}$,
$P^{(2)}_{\rm p}$, and $\xi^{(2)}_{\rm p}$ are summarized in Appendix~\ref{HigherPole}.

\subsection{Comparison between real and redshift space}
\label{Sec.5.1}

\begin{figure}[t]
	  \begin{center}
	  \psfig{figure=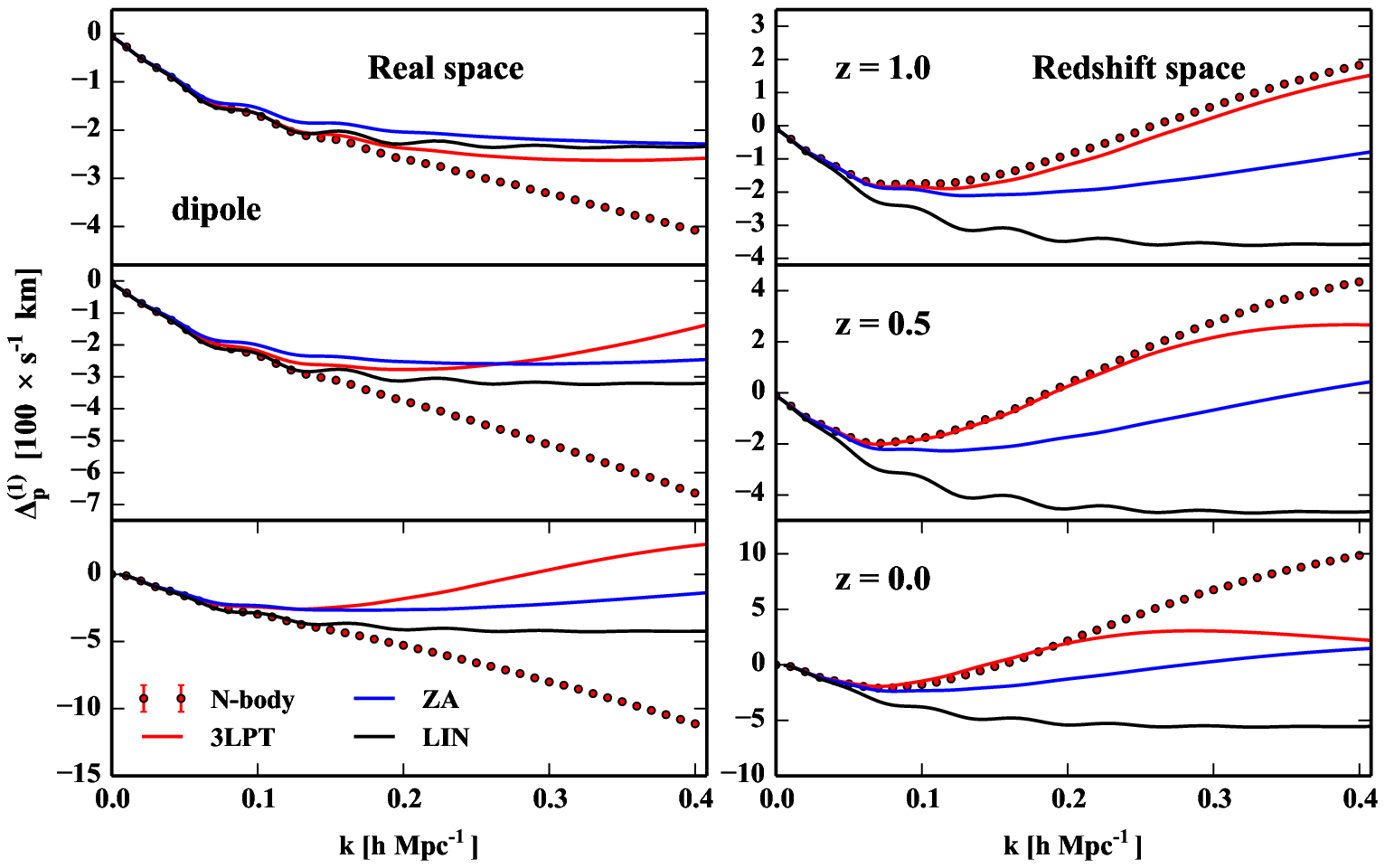}
		\end{center}
		\vspace{-0.5cm}
		\caption{ Dipole terms of the density-weighted velocity power spectra $\Delta^{(1)}_{\rm p}$
		defined in Eqs~(\ref{pk_red}) and (\ref{Delta})
		are plotted at redshifts of $z=0.0$, $0.5$, and $1.0$. 
		Each line denotes 3LPT (red), ZA (blue), and linearized theory (black),
		where 3LPT is the main result in this paper.
    		The $N$-body results are shown for dark matter particles (red points).
		Higher order perturbation corrections (3LPT), which is computed without using any free parameters,
		improve the Zel'dovich approximation until $k\sim 0.1\ h{\rm Mpc}^{^-1}$
   		at any redshift in both real and redshift space. }
		\label{fig2}
\end{figure}

To clarify the importance of redshift space distortions in the density-weighted velocity statistics,
we plot the dipole terms of $\Delta_{\rm p}^{(1)}$ and $\xi_{\rm p}^{(1)}$
in both real space and redshift space at the redshift of $z=0.0$ in Figure~\ref{fig1}.
Blue and red symbols denote the predictions from the $N$-body simulations in real space and redshift space, respectively.
Furthermore, we compute the results in the Zel'dovich approximation in two ways.
The first is the measurement from particle distributions in the Zel'dovich approximation (cyan and magenta symbols),
and the second is the numerically computed solution in the perturbation theory in the Zel'dovich approximation discussed in Sec.~\ref{TheoreticalModel}
(cyan and magenta solid lines).
If two objects (dark matter particles, halos, galaxies, and galaxy clusters) are moving toward each other,
their contribution to $\Delta_{\ell=1}^{(1)}$ and $\xi_{\ell=1}^{(1)}$ will be negative, $\Delta_{\ell=1}^{(1)}<0$ and $\xi_{\ell=1}^{(1)}<0$,
and if moving apart, positive $\Delta_{\ell=1}^{(1)}>0$ and $\xi_{\ell=1}^{(1)}>0$.
Gravitational attraction predicts a slight tendency of any pair of objects 
to be moving toward rather than away from each other at large scales, 
resulting in the negative values of $\Delta_{\ell=1}^{(1)}$ and $\xi_{\ell=1}^{(1)}$.

In real space, we find the change in sign of $\Delta_{\ell=1}^{(1)}$ from negative to positive in the Zel'dovich approximation 
at small scales around $k\sim 0.6\ h{\rm Mpc}^{-1}$ (see cyan symbols in the left panel of Figure~\ref{fig1}).
This means that the Zel'dovich particles move apart from each other at those scales,
because they do not form halos and pass through each other at small scales, smearing small scale structure.
On the other hand, the $N$-body simulation can form halos, and dark matter particles are trapped in halos,
resulting in the infall velocity until scales around virial radius of halos and random motion of particles within virial radius.
As the result, the $\Delta^{(1)}_{\ell=1}$ and $\xi^{(1)}_{\ell=1}$ measured from the $N$-body simulation keep to be negative 
within the range of scales in Figure~\ref{fig1} (see blue symbols).

In redshift space, the signs of $\Delta_{\ell=1}^{(1)}$ and $\xi_{\ell=1}^{(1)}$ change from negative to positive
in both the $N$-body simulation and the Zel'dovich approximation
at mildly non-linear scales ($k\sim 0.1\mathchar`- 0.3\ h{\rm Mpc}^{-1}$)~\cite{Okumura:2013zva}.
This is due to redshift space distortions which is the coordinate transformation from real to redshift space in Eq.~(\ref{Coord}).
Since the positions of particles move in the direction of their line-of-sight velocities through the coordinate transformation,
different two particles apparently path through and move away from each other in redshift space at mildly small scales.
It should be noted that even in the Zel'dovich approximation this effect appear,
and the sign of the Zel'dovich power spectrum changes at weakly non-linear scales
(from $k\sim 0.6\ h{\rm Mpc}^{-1}$ in real space to $k\sim 0.3\ h{\rm Mpc}^{-1}$ in redshift space). 
Thus, redshift space distortions give rive to the change in signs of $\Delta_{\ell=1}^{(1)}$ and $\xi_{\ell=1}^{(1)}$ at the mildly small scales
regardless of whether halos are formed in particle distributions.
In this sense, 
the change in signs occurs due to the non-linear velocity dispersion effect.
The so-called ``Finger-of-God (FOG)'' effect is a part of the effect,
because the FOG effect is the coordinate transformation by random motion of particles within halos.

Finally, in the Zel'dovich approximation 
the numerically computed solutions of Eqs.~(\ref{pk_red}) and (\ref{pk_an})
agree well with the measurements from particle distributions in the Zel'dovich approximation
for both the power spectrum and correlation function and in both real and redshift space.
This fact guarantees the validity of the technique to compute the non-linear power spectrum in the Lagrangian description 
with redshift space distortions discussed in Sec.~\ref{TheoreticalModel} (see also Appendix~\ref{ZA}).
Furthermore, this good agreement also implies the validity of an algorithm used to measure 
the density-weighted velocity statistics from simulation data,
because these two methods to compute the Zel'dovich power spectrum are independent of each other.
In the next subsection, we investigate how higher order perturbation corrections improve the Zel'dovich approximation.

\subsection{Comparison between perturbation theories and $N$-body simulations}

\begin{figure}[t]
		\begin{center}
	  \psfig{figure=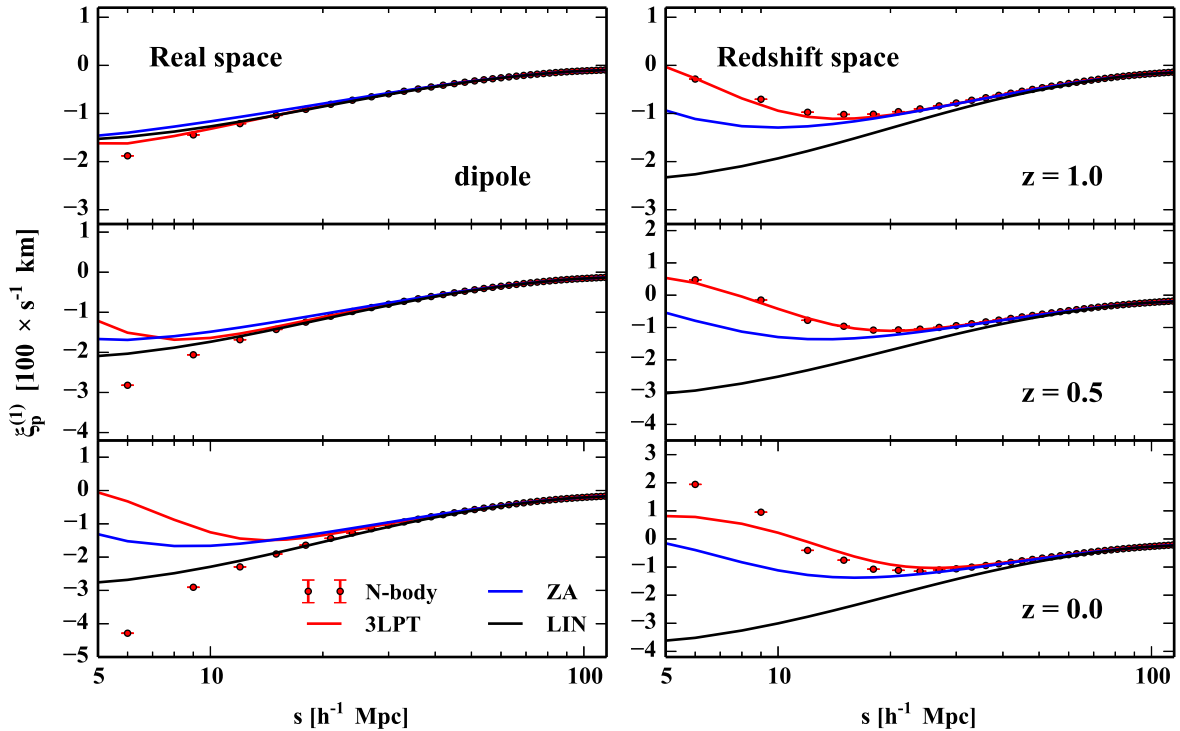}
		\end{center}
		\caption{ Dipole terms of the density-weighted velocity correlation function $\xi^{(1)}_{\rm p}$
		are plotted at redshifts of $z=0.0$, $0.5$, and $1.0$. 
		Each line denotes 3LPT (red), ZA (blue), and linearized theory (black),
		and the $N$-body results are shown for dark matter particles (red points).
		The quantity that was measured in~\cite{Hand:2012ui} 
		is the two-point correlation function of the galaxy halos which is closely 
		related to the dipole term of the matter two-point correlation function $\xi^{(1)}_{\rm p}$ in redshift space.
	      }
		\label{fig3}
\end{figure}

In Figures~\ref{fig2} and \ref{fig3},
we show $\Delta^{(1)}_{\ell=1}$ and $\xi^{(1)}_{\ell=1}$
computed using the perturbation theory discussed in Sec.~\ref{TheoreticalModel} and the $N$-body simulation at the redshifts of $z=0.0$, $0.5$, and $1.0$
and in both real and redshift space.

Our analytical approach in Sec.~\ref{TheoreticalModel}
can take account into non-linear effects from the coordinate transformation from real space to redshift space.
Therefore, we can interpret the main difference among the Zel'dovich approximation (blue line), the 3LPT solution (red line),
and the $N$-body simulation (red points) as non-linear gravitational effects which 
yields non-linear corrections to the displacement vector and peculiar velocity of particles.
While the Zel'dovich approximation and the 3LPT solution include the linear and third order gravitational potentials
in the perturbation expansion, the $N$-body simulation considers the full non-linear gravitational potential.

In real space, until scales around $k\sim 0.1\ h{\rm Mpc}^{-1}$ and $s\sim 30\ h^{-1}{\rm Mpc}$,
the 3LPT solution is a clear improvement over the Zel'dovich approximation and
is a better fit to the $N$-body simulations at any redshifts.
At small scales, the 3LPT solution emphasizes the feature of the Zel'dovich approximation that particles move away from each other,
because non-linear displacement vectors in 3LPT do not form halos.

In redshift space, 
one of the most important features of $\Delta_{\ell=1}^{(1)}$ and $\xi_{\ell=1}^{(1)}$ is the change in signs
of $\Delta_{\ell=1}^{(1)}$ and $\xi_{\ell=1}^{(1)}$ themselves at mildly small scales 
due to redshift space distortions as discussed in Sec.~\ref{Sec.5.1}
(see also~\cite{Okumura:2013zva}).
This characteristic feature is explained even by the Zel'dovich approximation.
This fact implies the importance of the full non-linearities from the continuity equation
for predicting the redshift space distortion in Sec.~\ref{TheoreticalModel},
because the Zel'dovich approximation only includes the linearized gravitational potential
as well as linear theory which is scale-independent.
Similarly, there is an obvious difference 
between the 3rd order standard perturbation theory (3SPT) and 3LPT as shown in Appendix~\ref{SPT},
where both of the theories include the 3rd order gravitational potential in the perturbation theory.
Comparing with the Zel'dovich approximation, 3LPT, and the $N$-body simulation,
we find that the change of the direction of the infall velocity is emphasized by the non-linear gravitational effect.
Compared to the results in real space,
we find that the 3LPT solution improves the Zel'dovich approximation at any redshifts
until $k\sim 0.1\ h{\rm Mpc}^{-1}$ and $s\sim 30\ h^{-1}{\rm Mpc}$ also in redshift space.

\begin{figure}[t]
	  \begin{center}
	  \psfig{figure=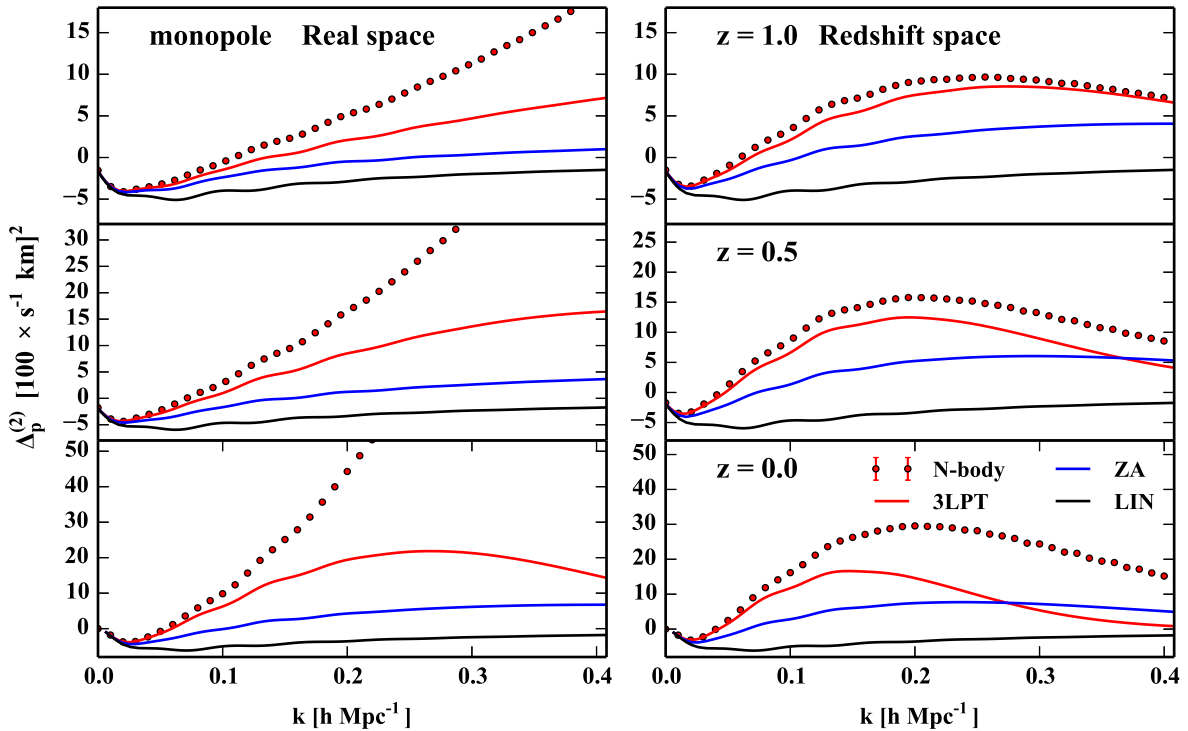}
		\end{center}
		\vspace{-0.5cm}
		\caption{
		Same as Figures~\ref{fig2}.
		These figures show the monopole terms of $\Delta^{(2)}_{\rm p}$.
    		}
		\label{fig4}
\end{figure}

\begin{figure}[t]
	  \begin{center}
	  \psfig{figure=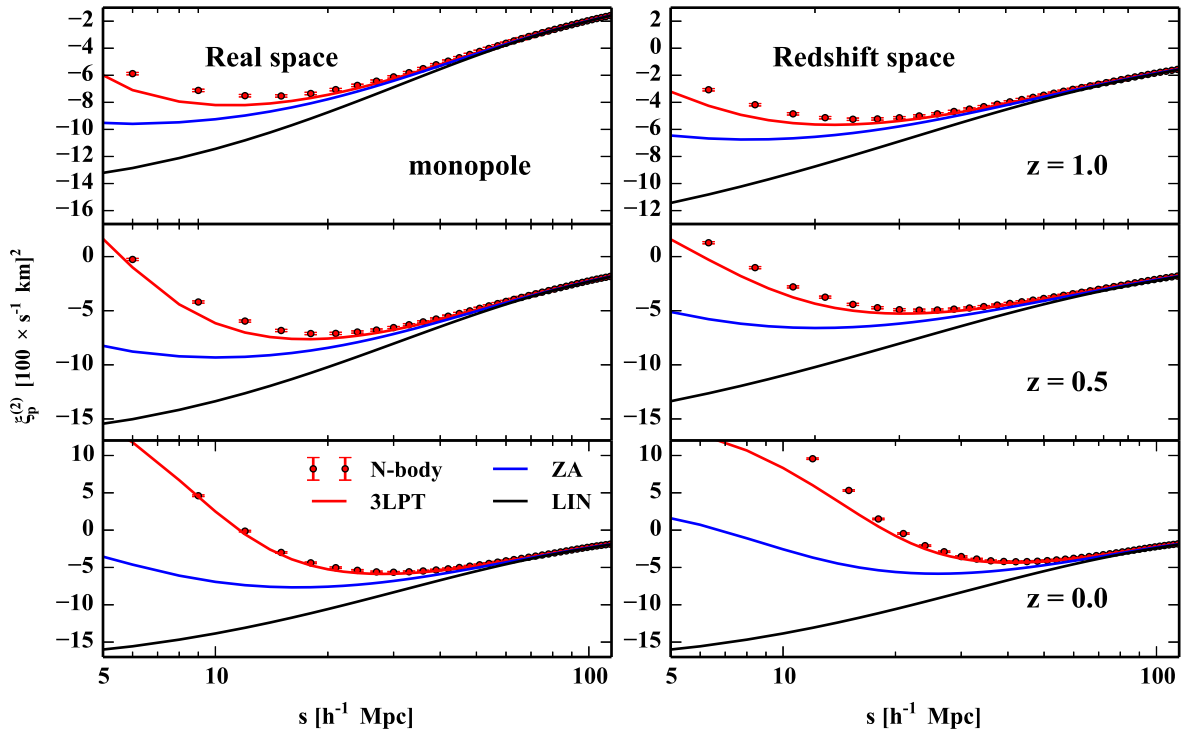}
		\end{center}
		\vspace{-0.5cm}
		\caption{
		Same as Figures~\ref{fig3}.
		These figures show the monopole terms of $\xi^{(2)}_{\rm p}$.
    		}
		\label{fig5}
\end{figure}

As another application of our formalism,
Figures~\ref{fig4} and \ref{fig5} show the monopole terms of $\Delta^{(2)}_{\rm p}$ and $\xi^{(2)}_{\rm p}$.
We find that the 3LPT solution is a better fit to the $N$-body simulations than the Zel'dovich approximation.
The 3LPT solutions can explain the $N$-body results until
$k\sim 0.05\ h{\rm Mpc}^{-1}$ and $s\sim 30\ h^{-1}{\rm Mpc}$ in both real and redshift space at any redshifts.

Formulation of the density-weighted velocity statistics in redshift space has been done in \cite{Okumura:2013zva} 
based on the distribution function approach (DFA) \cite{Seljak:2011tx}. 
\cite{Okumura:2013zva} adopted the Eulerian description and considered finite mode-coupling integrals corresponding to the 1-loop SPT. 
Because the PT model predictions diverge at high $k$, \cite{Okumura:2013zva} needed to introduce the smoothing function
to obtain the correlation function although the filter effects were shown to be negligible on scales larger than $5\ h^{-1}{\rm Mpc}$ \cite{Vlah:2013lia}. 
Furthermore, \cite{Okumura:2013zva} introduced free parameters for the nonlinear velocity dispersion for dark matter power spectrum,
based on the halo model \cite{Vlah:2012ni}.
On the other hand, our formalism use the Lagrangian description with infinite mode-couping integrals of wavenumber in computing the power spectra, 
and we do not need to introduce a free parameter for the nonlinear velocity dispersion and a window function to obtain the correlation function. 

The model predictions for the velocity correlation function were compared to $N$-body simulations in both real space 
\cite{Bhattacharya:2008,Reid:2011,Okumura:2013zva} and in redshift space \cite{Okumura:2013zva}, and the predictions were systematically smaller than the $N$-body measurements even at very large scales. This discrepancy could be explained by the finite volume effect. We take into account the finite volume effect in computing the correlation function, enabling to fit the analytical predictions to the $N$-body simulation results at linear regions (Appendix~\ref{Volume}).

\section{Conclusion}

This paper presents analytic predictions for the galaxy-weight KSZ signal, a measurement that combines
CMB observations with galaxy redshift surveys.
This signal measures the galaxy momentum in redshift space and is potentially a powerful
new cosmological probe. 

We derived a simple relation between the density power spectrum and the density-weighted velocity power spectrum
Eq.~(\ref{pk_red}) that holds for dark matter particles, the electron density field,  and halos.
In this paper, we use Lagrangian perturbation techniques to predict the non-linear matter power spectrum
and then compute the density-weighted velocity power spectrum.   If we had used alternative approaches
to compute the density power spectrum, Eq.~(\ref{pk_red}) could still be used
to compute the density-weight power spectrum.

Using  third order  LPT, we compute the predicted galaxy-weighted velocity field
and compare with the results of $N$-body simulations (Figures~\ref{fig2} and~\ref{fig3}).
While we  use LPT to approximately treat the non-linear gravitational effects, we follow 
~\cite{Carlson:2012bu,Sugiyama:2013mpa,Vlah:2014nta} and do not 
make any approximations in our Lagrangian calculation of the density non-linearities.
Rather, we use the continuity equation and represent the transformation to redshift space as
a coordinate transformation.  
In the mildly non-linear regions,  scales larger than $\simeq 30\ h^{-1}{\rm Mpc}$, 
our predictions agree with the simulation at various redshifts of $z=0.0$, $z =0.5$, and $z=1.0$ in both real and redshift space.
Our model explains the change in sign in the KSZ two point correlation function at small scales.

This effect is due to the change in the direction of infall velocity of particles in redshift space.
This characteristic feature appears even in the Zel'dovich approximation
that only includes the linearized gravitational potential,
and is enhanced when we include  non-linear gravitational effects through higher order perturbation theory.
Our prediction can explain this feature without using any free parameters: this underscores the importance of considering the full non-linear effect by describing redshift space
distortions as a  coordinate transformation.

In this paper, we have calculated the non-linear matter density field.  The next
step in the calculation is compute the non-linear halo density field.
This requires computing the 
relation between dark matter distribution to halo distribution.
 Appendix~\ref{Halos} shows significant differences 
 between dark matter particles and halos even on mildly non-linear scales.
 Thus, it will be necessary to accurately 
model  halo bias even at scales where the perturbation theory works well.
We defer 
modeling the scale-dependent halo bias to a future paper.

\acknowledgments

We thank E.Schaan, M. Shirasaki, K. Osato, and N. Yoshida for useful comments.
Numerical computations were carried out on Cray XC30
at Center for Computational Astrophysics, National Astronomical Observatory of Japan.
NSS was supported by a grant from the Japan Society for the Promotion of Science (JSPS) (No. 24-3849) during his stay at Princeton.
NSS acknowledges financial support from Grant-in-Aid for Scientific Research from the JSPS Promotion of Science (25287050).
T.O. was supported by Grant-in-Aid for Young Scientists (Start-up) from the Japan Society for the Promotion of Science (JSPS) (No. 26887012).
D.N.S. was partially supported by NSF  grant AST-1311756 and NASA grants NNX12AG72G and NNX14AH67G.

\providecommand{\href}[2]{#2}\begingroup\raggedright\endgroup

\appendix

\section{Baryon acoustic peak}
\label{BAO}

\begin{figure}[t]
		\begin{center}
	  	\psfig{figure=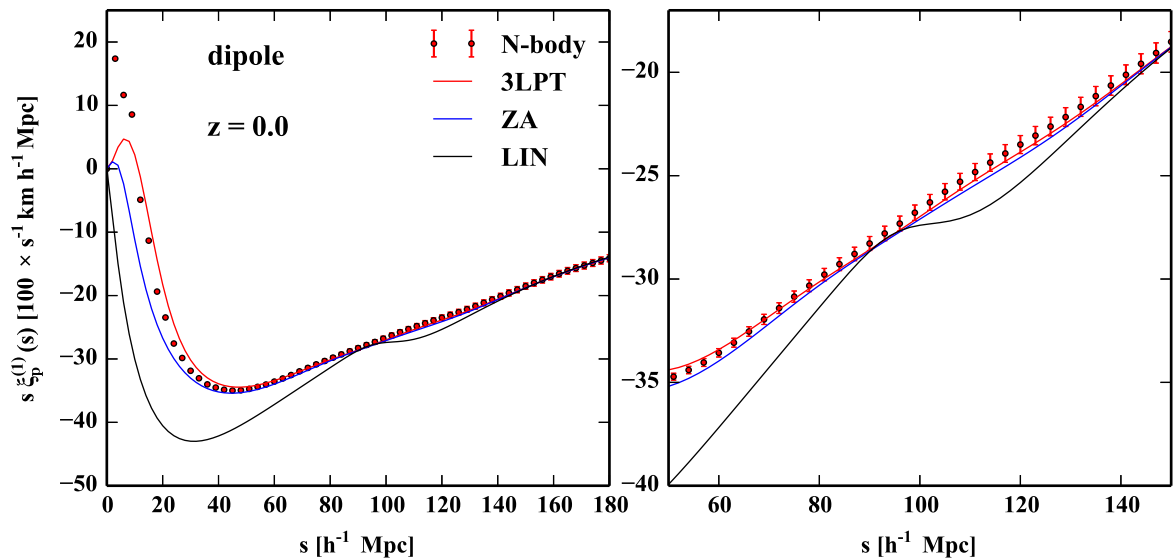}
		\end{center}
		\vspace{-0.5cm}
		\caption{
		These figures show the dipole term of $s \xi_{\rm p}^{(1)}(s)$ at a redshift of $z=0.0$.
		The right panel is the same as the left panel, but it is shown in the range of $50h^{-1}{\rm Mpc} < s < 150h^{-1}{\rm Mpc}$.
		While the baryon acoustic peak is visible in the linearized theory
		around the scales of $\sim100h^{-1}{\rm Mpc}$,
		the peak is broadened in the other non-linear predictions.
		In particular, the Zel'dovich approximation can explain the smearing of the baryon acoustic peak.
		}
		\label{fig:BAO}
\end{figure}

Figure~\ref{fig:BAO} shows $s \xi_{\ell=1}^{(1)}(s)$,
which corresponds to the cross-correlation between the density-weighted velocity along the line-of-sight and the density field,
in the linearized theory (black), the Zel'dovich approximation (blue),
3LPT (red), and the $N$-body simulation (red symbols) at a redshift of $z=0$.
While the baryon acoustic peak is visible in the linearized theory around the scales of $100h^{-1}{\rm Mpc}$,
the peak broadens in the other non-linear predictions.
In particular, we find that the Zel'dovich approximation can explain the smearing of the baryon acoustic peak.

\section{Higher-pole terms}
\label{HigherPole}

In this appendix, we investigate higher pole terms of both the power spectrum and two-point correlation function in the Legendre expansion,
which are generated by anisotropies from redshift space distortions.

\begin{figure}[t]
		\begin{center}
	  	\psfig{figure=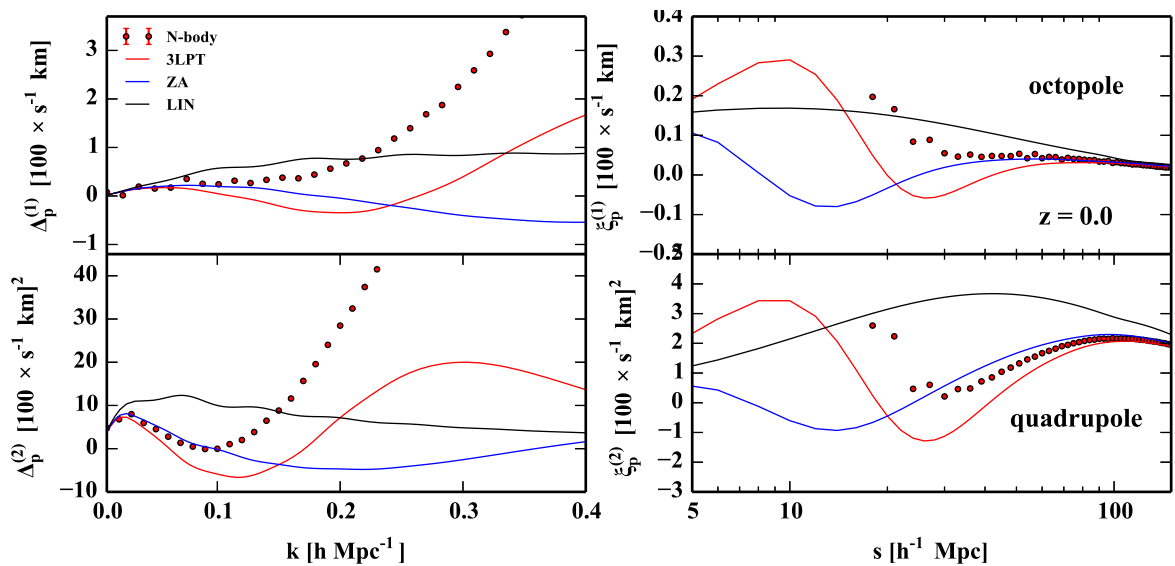}
		\end{center}
		\caption{
		Higher-pole (octopole and quadrupole) terms of $\Delta^{(1)}_{\rm p}$, $\xi^{(1)}_{\rm p}$,
		$\Delta^{(2)}_{\rm p}$, and $\xi^{(2)}_{\rm p}$ are plotted at a redshift of $z=0.0$.
		In the two-point correlation functions,
		even at large scales of $> 60\ [{\rm Mpc}/h]$ the non-linear effects appear.
		Therefore, the Zel'dovich approximation or the 3LPT solution,
		not the linearized theory, are needed for explaining the $N$-body simulation results.
    		}
		\label{fig:Multi}
\end{figure}

In Figure~\ref{fig:Multi}, we plot the octopole terms of $\Delta^{(1)}_{\rm p}$ and $\xi^{(1)}_{\rm p}$
and the quadrupole terms of $\Delta^{(2)}_{\rm p}$ and $\xi^{(2)}_{\rm p}$, respectively.
In particular, redshift space distortions yield the octopole terms of $\Delta_{\rm p}^{(1)}$ and $\xi_{\rm p}^{(1)}$,
while the quadrupole terms of $\Delta_{\rm p}^{(2)}$ and $\xi_{\rm p}^{(2)}$ appear also in real space.
Figure~\ref{fig:Multi} shows that
the analytical predictions do not work very well at small scales and at the redshift of $z=0.0$.
However, note that even at large scales of $> 60\ [{\rm Mpc}/h]$, the non-linear effects become important
in the two-point correlation functions (lower panels).
As a result, we find that for explaining the $N$-body simulation results,
we need the Zel'dovich approximation or the 3LPT solution, not the linearized theory.
We have checked that the 3LPT solution works better at higher redshifts, $z=0.5$ and $1.0$.

\section{Zel'dovich approximation}
\label{ZA}

We have presented the analytical form for the power spectrum using the displacement vector in redshift space as shown in equation (\ref{pk_an}). 
In the situation that we have perfect knowledge about the particle distribution, 
namely we can compute $\YY_s$ for each particle,
we expect to obtain both the same power spectrum and correlation function from theory and simulations. 
The Zel'dovich approximation is an ideal example for this purpose, because linearized perturbation theory gives an exact value for the displacement vector.

\begin{figure}[p]
		\begin{center}
	  	\psfig{figure=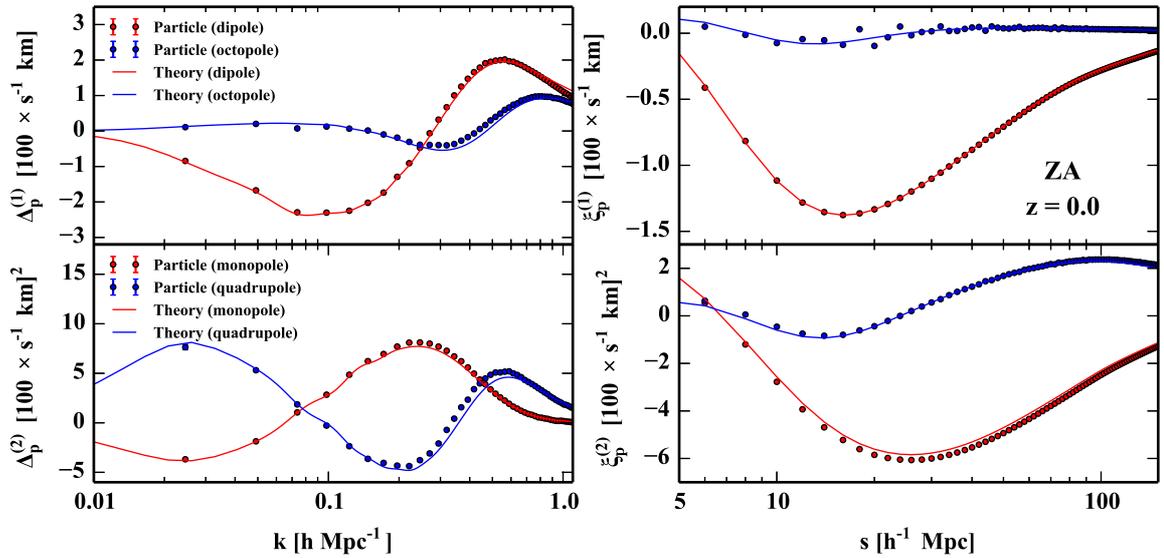}
		\end{center}
		\caption{
		Two predictions from the analytical calculation in Eqs.~(\ref{Pkan}) and (\ref{pk_red})
		and the measurement from the Zel'dovich particle distribution which is generated by the $2LPT$ code~\cite{Crocce:2006ve} are shown.
		The dipole, octopole, monopole, and quadrupole terms of $\xi_{\rm p}^{(1)}$, $\Delta_{\rm p}^{(1)}$,
		$\xi_{\rm p}^{(2)}$, and $\Delta^{(2)}_{\rm p}$ are plotted at a redshift of $z=0.0$.
    		}
		\label{fig:ZA}
\end{figure}

Figure~\ref{fig:ZA} shows the two predictions from the analytical calculation in Eqs.~(\ref{Pkan}) and (\ref{pk_red})
and the measurement from the Zel'dovich particle distribution which is generated by the $2LPT$ code~\cite{Crocce:2006ve}.
At the redshift of $z=0$, our predictions can explain the results from the particle distribution well.
This fact ensures that the approximation method in Eq.~(\ref{Pkan}) for computing 
the power spectrum with works well.
The discrepancy between the results from theory and simulations at small scales
may come from the truncation of the expansion of the power spectrum in Eq.~(\ref{Pkan}) at the third order $n=2$.
Otherwise, it may be due to numerical errors in measuring the power spectrum and correlation function 
from a particle data in the Zel'dovich approximation.

\section{SPT vs. LPT}
\label{SPT}

The 3rd order standard perturbation theory is given by expanding the exponential factor
in the estimator in Eq.~(\ref{pk_an})
and truncating at ${\cal O}\left( P_0^2 \right)$~\cite{Crocce:2006,Sugiyama:2013mpa}
\begin{eqnarray}
	 P(\kk) =  \int d^3q e^{-i\kk\cdot\qq}\left\{  e^{\Sigma(\kk,\qq)-\bar{\Sigma}(\kk)}  \right\}
	  = \sum_{n=0}^{\infty}\frac{1}{n!}\int d^3q e^{-i\kk\cdot\qq} 
	  \left(   {\Sigma(\kk,\qq)-\bar{\Sigma}(\kk)} \right)^n
	  + {\cal O}\left( P_0^3 \right),
\end{eqnarray}
where $P_{0}$ is the linearized power spectrum.
In this sense, the difference from 3LPT computed in this paper 
is whether the estimator is kept or not.
Note that the gravitational potential in both of 3LPT and 3SPT 
is the same: namely, the third order in the perturbation expansion.

\begin{figure}[p]
		\begin{center}
	  		\psfig{figure=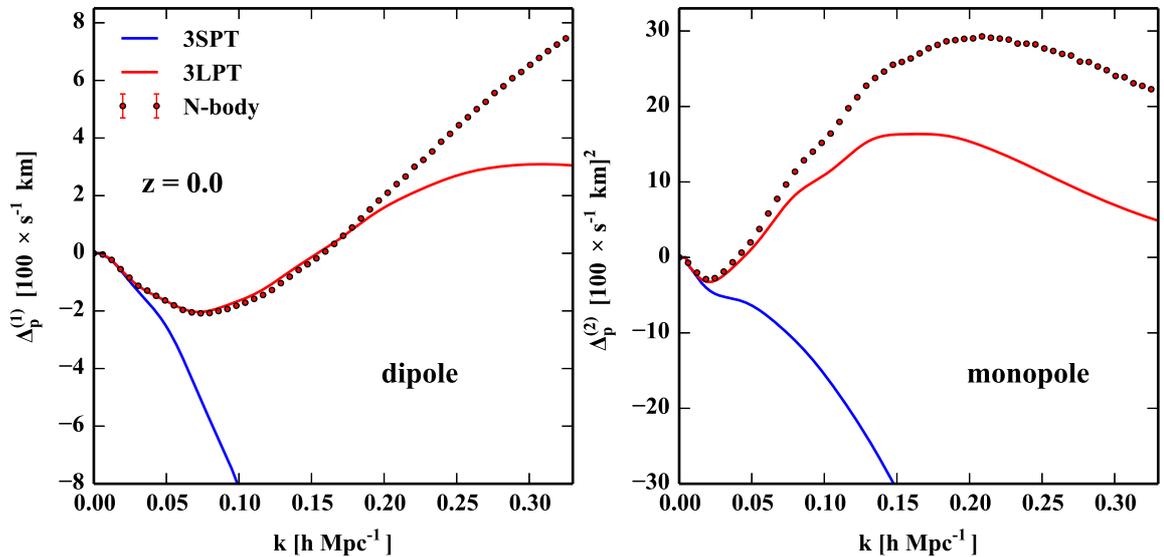}
		\end{center}
		\caption{
		Comparison between the 3rd order standard perturbation theory (3SPT, blue)
		and Lagrangian perturbation theory (3LPT, red) is shown at a redshift of $z=0.0$.
		Although both of these theories include the 3rd order gravitational potential in 
		the perturbation expansion, there is the obvious difference between them.
		These figures imply the success of our approximation method
		to compute the power spectrum in Eq.~(\ref{Pkan}).
    		}
		\label{fig:SPT}
\end{figure}

Figure~\ref{fig:SPT} compares the 3SPT and 3LPT solutions of $\Delta_{\rm p,\ell}^{(1)}$ and $\Delta_{\rm p,\ell}^{(2)}$
with the redshift space distortion at a redshift of $z=0.0$.
Clearly, the 3SPT solutions do not explain the $N$-body simulation results,
while the 3LPT solutions works well.
This fact implies the success of our approximation method to compute the power spectrum in Eq.~(\ref{Pkan}).

\section{Finite volume effects}
\label{Volume}

\begin{figure}[t]
		\begin{center}
	  		\psfig{figure=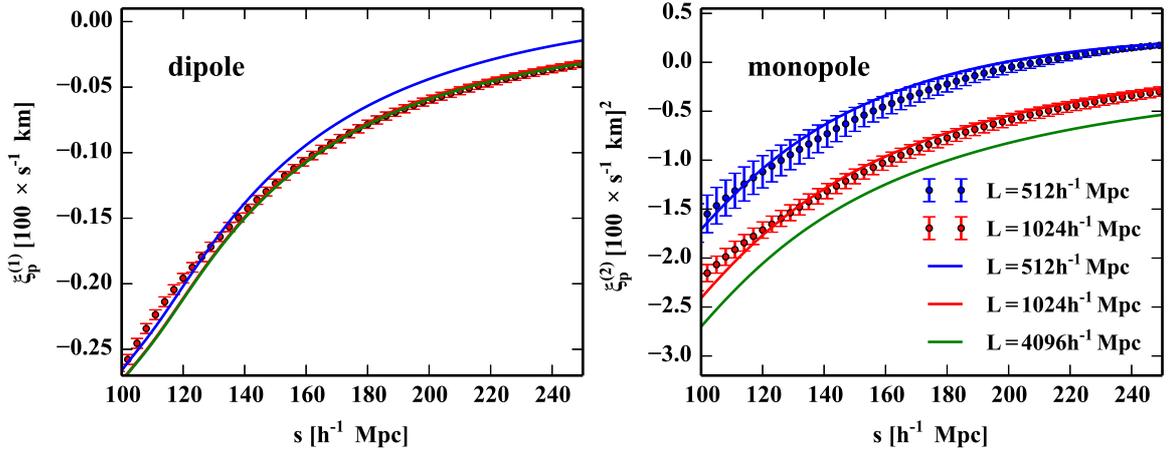}
		\end{center}
		\vspace{-0.5cm}
		\caption{
			  Density-weighted correlation functions, $\xi_{\rm p}^{(1)}$ and $\xi_{\rm p}^{(2)}$, are plotted at a redshift of $z=0.0$
		with different volumes.
		Red and blue symbols show the predictions from $N$-body simulations with box sizes of $L = 512\ {\rm and}\ 1024h^{-1}{\rm Mpc}$,
		and solid lines are predictions in linear theory. 
		The minimum wavenumber in inverse Fourier transform to compute the correlation functions in linear theory
		is determined by $k_{\rm min} = 2.0\pi/L$, where $L$=512, 1024, and $4096h^{-1}{\rm Mpc}$.
		For display purposes, the dipole term from the $N$-body simulation with $L = 512h^{-1}{\rm Mpc}$ is not shown, because it is noisy.
    		}
		\label{fig:Volume}
\end{figure}

Figure~\ref{fig:Volume}
shows the impact of changing the minimum wavenumber $k_{\rm min} = 2\pi/L$ to compute 
the density-weighted velocity correlation functions $\xi_{\rm p}^{(1)}$ and $\xi_{\rm p}^{(2)}$ in inverse Fourier transform
from the density-weighted velocity power spectra $P_{\rm p}^{(1)}$ and $P_{\rm p}^{(2)}$.
In computing $\xi^{(1)}_{\rm p}$ and $\xi^{(2)}_{\rm p}$ in linear theory,
we use the minimum wavenumber corresponding to the box size of the simulation in inverse Foueir transform
$k_{\rm min}=2\pi/L$, where $L^3$ is a simulation volume.
we predict different amplitudes of the correlation function from the $N$-body simulation results even in linear theory at very large scales.

\section{Halos}
\label{Halos}

Since the density-weighted velocity is well-defined measurable quantity,
we can measure the power spectra and the two-point correlation functions of the density-weighted velocity for halos.
Here, we identify halos using the friend of friend (FOF) method.
We considered three kinds of mass range:
$1.0\times 10^{13}< M < 5.0\times 10^{13}M_{\odot}$,
$5.0\times 10^{13}< M < 1.0\times 10^{14}M_{\odot}$,
and $1.0\times 10^{14}M_{\odot}< M$.

\begin{figure}[h]
		\begin{center}
	  		\psfig{figure=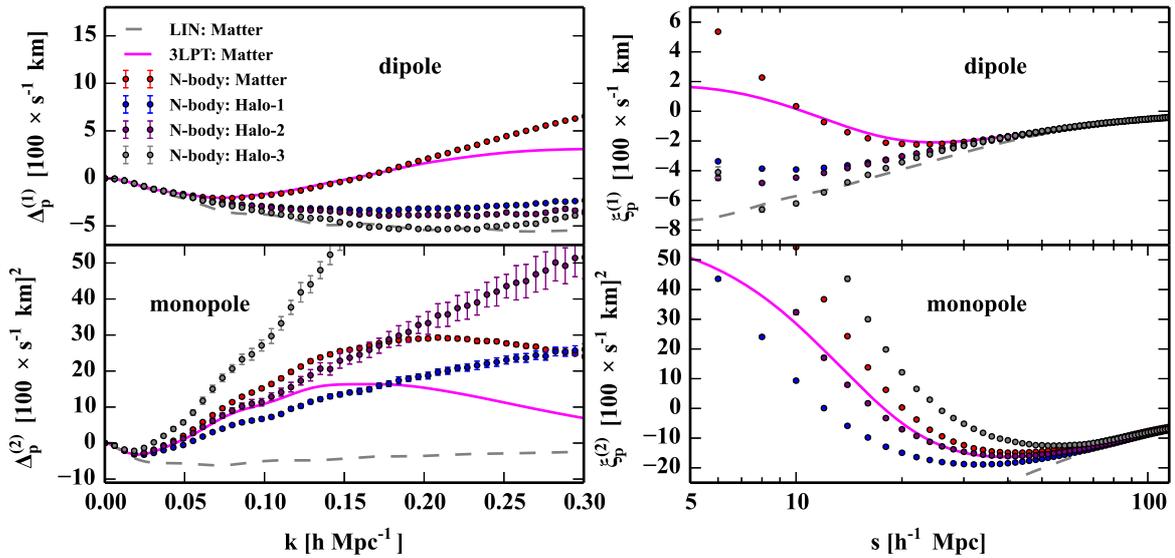}
		\end{center}
		\caption{
		The dipole and monopole terms of
		$\Delta_{\rm p}^{(1)}$, $\xi_{\rm p}^{(1)}$, $\Delta_{\rm p}^{(2)}$, and $\Delta_{\rm p}^{(2)}$
		for dark matter and halos are plotted at a redshift of $z=0.0$.
		Three mass ranges of halos are chosen:
		$1.0\times 10^{13}\ M_{\odot} < M < 5.0\times 10^{13}\ M_{\odot}$
		(Halo-1, blue points),
		$5.0\times 10^{13}\ M_{\odot} < M < 1.0\times 10^{14}\ M_{\odot}$
		(Halo-2, purple points),
		and $1.0\times 10^{14}\ M_{\odot} < M$ (Halo-3, gray points).
		The dipole terms of $\Delta_{\rm p}^{(1)}$ and $\xi_{\rm p}^{(1)}$ for halos
		are normalized by the linear Kaiser factor $\left( 1 + 3f/5\right)/\left( b + 3f/5 \right)$
		with $b$ being the linear spatial bias,
		so that they agree with $\Delta_{\rm p}^{(1)}$ and $\xi_{\rm p}^{(1)}$ for dark matter at large scales.
		Since the velocity bias converges to unity at large scales,
		$\Delta_{\rm p}^{(2)}$ and $\xi_{\rm p}^{(2)}$ for halos and dark matter become similar to each other at large scales.
		Therefore, these figures show the scale-dependent effects of spatial and velocity bias.
    		}
		\label{fig:Halos}
\end{figure}

In Figure~\ref{fig:Halos},
we plotted the dipole and monopole terms 
of $\Delta_{\rm p}^{(1)}$, $\xi_{\rm p}^{(1)}$, $\Delta_{\rm p}^{(2)}$, and $\Delta_{\rm p}^{(2)}$
for dark matter particles and halos at a redshift of $z=0.0$.
We normalized the dipole terms of $\Delta_{\rm p}^{(1)}$ and $\xi_{\rm p}^{(1)}$
using the linear Kaiser factor 
$ \left( 1 + 3 f / 5 \right) / \left( b + 3 f / 5 \right)$ with $b$ being the linear spatial bias,
so that the amplitudes of $\Delta_{\rm p}^{(1)}$ and $\xi_{\rm p}^{(1)}$ for halos
agree with those of dark matter particles at large scales.
Since the velocity bias is small to be ignored at large scales,
$\Delta_{\rm p}^{(2)}$ and $\xi_{\rm p}^{(2)}$ for dark matter particles and halos
become similar to each other at large scales.
As a result, Figure~\ref{fig:Halos} shows the scale-dependence of the spatial and velocity bias.
The contributions from the scale-dependent bias
become important at less scales than $\sim 30\ h^{-1}{\rm Mpc}$ at a redshift of $z=0.0$.
We can find the change in the direction of the infall velocity along the line of sight also for halos
as well as dark matter particles.

\newpage


\end{document}